\documentclass[aps,pra,showpacs,twocolumn,superscriptaddress]{revtex4}
\usepackage{amssymb}
\usepackage{amsmath}
\usepackage{graphicx}
\usepackage{epsfig}
\usepackage{subfigure}

\begin{document}

\title{Optical unidirectional amplification in a three-mode optomechanical
system}
\author{Y. Li}
\email{liyong@csrc.ac.cn}
\affiliation{Beijing Computational Science Research Center, Beijing, 100193, China}
\affiliation{Synergetic Innovation Center for Quantum Effects and Applications,
Hunan Normal University, Changsha, Hunan, 410081, China}
\author{Y. Y. Huang}
\affiliation{Beijing Computational Science Research Center, Beijing, 100193, China}
\author{X. Z. Zhang}
\affiliation{Beijing Computational Science Research Center, Beijing, 100193, China}
\affiliation{College of Physics and Materials Science, Tianjin Normal University, Tianjin
300387, China}
\author{L. Tian}
\affiliation{University of California, Merced, 5200 North Lake Road, Merced, California
95343, USA}

\begin{abstract}
We study the directional amplification of an optical probe field in a
three-mode optomechanical system, where the mechanical resonator interacts
with two linearly-coupled optical cavities and the cavities are driven by
strong optical pump fields. The optical probe field is injected into one of
the cavity modes, and at the same time, the mechanical resonator is subject to a mechanical drive with the driving frequency equal to the frequency difference between the optical probe and pump fields.
We show that the transmission of the probe field can be amplified in one direction and de-amplified in the opposite direction. This directional amplification or de-amplification results from the constructive or destruction interference between different transmission paths in this three-mode optomechanical system.
\end{abstract}

\maketitle

\section{Introduction}

\label{Introduction}

With the rapid development of microfabrication technology, cavity
optomechanical system~\cite{Kippenberg,Aspelmeyer1,Meystre,Aspelmeyer3} is
becoming an appealing candidate to connect a broad spectrum of photonic, electronic, and atomic devices, besides being studied for fundamental
questions of macroscopic systems in the quantum limit~\cite{Vitali}.
Recently, enormous progresses have been achieved that aim at the
applications of optomechanical systems in ultra-high precision measurement ~%
\cite{Rugar,Krause,Regal,Teufel,Forstner,Xu,Arvanitaki}, quantum information
processing~\cite{Mancini}, quantum illumination~\cite{Barzanjeh} to
optomechanically induced transparency~\cite%
{OMIT1,OMIT2,OMIT3,OMIT4,OMIT5,OMslowlight1,OMslowlight2}, absorption~\cite%
{OMIA1,OMIA2}, and amplification ~\cite%
{OMAmplification1,OMAmplification2,OMAmplificationJia,OMAmplificationXu,OMAmplificationSSi}.

Among these applications, nonreciprocal transmission and amplification are
of great interest in the study of the quantum analogue of photonic and
electronic devices, such as diode, circulator, and
transistor, which are crucial for scalable quantum information processing
in integrated circuits~\cite{Haus}. In the past, nonreciprocal devices have
been investigated broadly in optical systems~\cite%
{Faraday,ParametricModulation1,ParametricModulation2,ParametricModulation3,ParametricModulation4,Chang,Tang,Chiral,Lecocq}%
. In these devices, the occurrence of nonreciprocal light propagation is
associated with the symmetry breaking induced by various mechanisms, such as
magneto-optical Faraday effect~\cite{Faraday}, parametric modulation~\cite%
{ParametricModulation1,ParametricModulation2,ParametricModulation3,ParametricModulation4}%
, optical nonlinearity~\cite{Chang,Tang}, and chiral light-matter
interaction~\cite{Chiral}.

In recent years, it has been shown that the optomechanical system can be
utilized to realize nonreciprocal effects for propagating light fields~\cite%
{Manipatruni,Hafezi,shen,Kim,Fang}.
The nonreciprocal optical diodes are achieved in multi-mode optomechanical systems with effective breaking of time-reversal symmetry generated by on-demand gauge-invariant phases~\cite{XXW,XXW2,Tian,Metelmann,YLZhang}.
Nonreciprocal phenomena with directional amplification have been
explored theoretically in general coupled-mode systems~\cite{Ranzani}. The phenomena of optical directional amplification have also been implemented experimentally very recently in multi-mode optomechanical systems~\cite{Ruesink,Mohammad,Bernier,Peterson,Malz}.

In this paper, we study a scheme to achieve directional amplification of
an optical probe field in a three-mode optomechanical system, where a
mechanical resonator is coupled to two optical modes that directly interact
with each other. In this system, controllable phase difference between the
linearized optomechanical couplings, which breaks the time-reversal symmetry
of this three-mode system, is generated by the strong pump fields on the
optical cavities. Meanwhile, the probe field is applied to one of the cavities and the mechanical resonator is subject to a mechanical drive with the driving frequency equal to the frequency difference between the optical probe and pump fields. The constructive (destructive) interference between the transmission paths for the optical probe field and its mechanical counterpart via the optomechanical interaction results in the amplification of the probe field~\cite{OMAmplificationJia}. Strong directional amplification of the optical field with high amplification ratio can be achieved in this system. In comparison with the previous works~\cite{Ruesink,Mohammad,Bernier,Peterson,Malz} in multi-mode optomechanical systems, where the directional amplification results from the blue-detuned pump fields, here we use the red-detuned pump fields as well as the additional mechanical drive to achieve the optical directional amplification in a three-mode optomechanical system. Since the blue-detuned (red-detuned) pump field will heat (cool) the motion of mechanical resonator in an optomechanical system, our scheme avoiding pumping with blue-detuned light can improve the stability of the amplification scheme in optomechanical systems. As a tradeoff, the additional mechanical drive with the driving frequency equal to the frequency difference between the optical probe and pump fields is required to achieve the directional amplification in our scheme. Our work provides an alternative method to achieve the optical directional amplification in optomechanical systems, which could stimulate future studies of optomechanical interfaces in the implementation of nonreciprocal and nonlinear photonic devices.

This paper is organized as follows. In Sec.~\ref{model}, we present the
Hamiltonian of the three-mode optomechanical system for nonreciprocal
amplification and our derivation of the transmission coefficients in this
system. Details of the directional amplification and de-amplification of
the optical probe field are studied in Sec.~\ref{transmission}. Conclusions
are given in Sec.~\ref{summary}.

\section{Model and transmission matrix}

\label{model}

\begin{figure}[tbp]
\centering
\includegraphics[bb=57 376 541 680, width=8.5cm, clip]{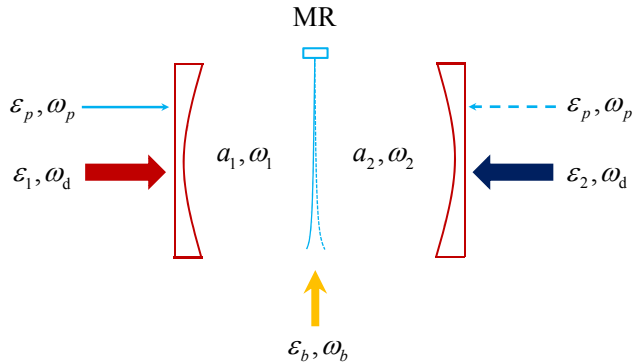} %
\caption{Schematic of a three-mode optomechanical system
driven by two pump fields with the same frequency $\omega _{d}$. A probe
field with frequency $\omega _{p}$ is applied to one of the two cavities,
that is, incident in cavity 1 from the left side (the thin solid arrow) or incident in cavity 2 from the right side (the thin dashed arrow).
The mechanical resonator is subject to a mechanical drive with the driving frequency $\omega_{b}$. The cavities and the mechanical resonator are coupled via radiation-pressure forces, and the cavities are directly coupled to each
other.}
\label{fig1}
\end{figure}

The optomechanical system under consideration consists of a mechanical
oscillator with resonance frequency $\omega _{m}$ and two optical cavities
with resonance frequencies $\omega _{1}$ and $\omega _{2}$, respectively, as
illustrated in Fig.~\ref{fig1}. We first focus on the case
that the probe field is incident from the left side to the cavity 1. The total
Hamiltonian of this system has the form
\begin{equation}
H=H_{0}+H_{I}+H_{d}.  \label{H}
\end{equation}%
The first term describes the free Hamiltonian of the cavity modes and the
mechanical one with ($\hbar =1$)
\begin{equation}
H_{0}=\omega _{1}a_{1}^{\dag }a_{1}+\omega _{2}a_{2}^{\dag }a_{2}+\omega
_{m}b^{\dag }b,
\end{equation}%
where $a_{i}^{\dag }$ ($a_{i}$) for $i=1,2$ and $b^{\dag }$ ($b$) are the
creation (annihilation) operators for the cavity modes and the mechanical one.
The second term
\begin{equation}
H_{I}=J\left( a_{1}^{\dag }a_{2}+a_{1}a_{2}^{\dag }\right)
+\sum_{i}g_{i}a_{i}^{\dag }a_{i}\left( b+b^{\dag }\right)
\end{equation}%
characterizes the linear coupling between the cavity modes with coupling
strength $J$ and the radiation-pressure force interaction between the cavities
and the mechanical resonator with single-photon coupling strength $g_{i}$. The
third term $H_{d}$ describes the mechanical drive, the optical pump fields on the cavities, and the probe field (incident from the left side to cavity 1, see the thin solid arrow in Fig.~\ref{fig1})
\begin{eqnarray}
H_{d} &=&\sum_{i}\left( i\varepsilon _{i}a_{i}^{\dag }e^{-i\omega
_{d}t}e^{i\theta _{i}}+\textrm{h.c.}\right)  \nonumber \\
&&+\left( i\varepsilon _{p}a_{1}^{\dag }e^{-i\omega _{p}t}+i\varepsilon
_{b}b^{\dag }e^{-i\omega _{b}t}+\textrm{h.c.}\right) ,
\end{eqnarray}%
where $\omega _{d}$ is the frequency, $\varepsilon _{i}$ is the amplitude,
and $\theta _{i}$ is the phase of the two pump fields, $\omega _{p}$ ($%
\omega _{b}$) is the frequency and $\varepsilon _{p}$ ($\varepsilon _{b}$)
is the amplitude of the probe field on cavity $1$ (the mechanical drive applied on the mechanical resonator). It is worth pointing out that the mechanical drive can be easily realized in experiments through an external electric drive~\cite{Rugar1,Okamoto,Faust,Fu}.
Here without loss of generality, we have assumed that $J$, $g_{1,2}$, and $%
\varepsilon _{1,2}$ are real numbers.

In the rotating frame with respect to the frequency of the pump fields, the quantum Langevin equations (QLEs) for the operators in the system are given by
\begin{widetext}
\begin{eqnarray}
\dot{a}_{1} &=&\left\{ -\gamma _{1}-i\left[ \Delta _{1}+g_{1}\left(
b+b^{\dag }\right) \right] \right\} a_{1}-iJa_{2} +\varepsilon
_{1}e^{i\theta _{{1}}}+\varepsilon _{p}e^{i(\omega _{d}-\omega _{p})t}+\xi
_{1}, \\
\dot{a}_{2} &=&\left\{ -\gamma _{2}-i\left[ \Delta _{2}+g_{2}\left(
b+b^{\dag }\right) \right] \right\} a_{2}-iJa_{1} +\varepsilon
_{2}e^{i\theta _{{2}}}+\xi _{2}, \\
\dot{b} &=&\left( -\gamma _{m}-i\omega _{m}\right) b-i\left(
g_{1}a_{1}^{\dag }a_{1}+g_{2}a_{2}^{\dag }a_{2}\right) +\varepsilon
_{b}e^{-i\omega _{b}t}+\xi _{m}.
\end{eqnarray}%
\end{widetext}
Here $\Delta _{i}=\omega _{i}-\omega _{d}$ ($i=1,2$) are the optical
detunings of the cavities, $\gamma _{i}$ ($\gamma _{m}$) are the decay rates
of the two cavities (mechanical resonator), $\xi _{i}$ ($\xi _{m}$) are the
noise operators of the cavities (mechanical mode) with $\langle \xi
_{i}\rangle =\langle \xi _{m}\rangle =0$.

We first derive the steady-state solution of the three-mode system under
strong pump fields. Neglecting the effects of the optical probe field and mechanical drive, we can obtain the steady-state solution
as
\begin{eqnarray}
\left\langle a_{1}\right\rangle & =& \frac{\left( \gamma _{2}+i\Delta
_{2}^{\prime }\right) \varepsilon _{1}e^{i\theta _{1}}-iJ\varepsilon
_{2}e^{i\theta _{2}}}{\left( \gamma _{1}+i\Delta _{1}^{\prime }\right)
\left( \gamma _{2}+i\Delta _{2}^{\prime }\right) +J^{2}},  \label{alpha1} \\
\left\langle a_{2}\right\rangle & =& \frac{\left( \gamma _{1}+i\Delta
_{1}^{\prime }\right) \varepsilon _{2}e^{i\theta _{2}}-iJ\varepsilon
_{1}e^{i\theta _{1}}}{\left( \gamma _{1}+i\Delta _{1}^{\prime }\right)
\left( \gamma _{2}+i\Delta _{2}^{\prime }\right) +J^{2}},  \label{alpha2} \\
\left\langle b\right\rangle & =& \frac{-i\left( g_{{1}}\left\vert
\left\langle a_{1}\right\rangle \right\vert ^{2}+g_{{2}}\left\vert
\left\langle a_{2}\right\rangle \right\vert ^{2}\right) }{\gamma
_{m}+i\omega _{m}},  \label{beta}
\end{eqnarray}%
where $\langle a_{i}\rangle$ ($\langle b\rangle$) are the steady-state
averages of the cavities (mechanical mode), and $\Delta _{i}^{\prime
}=\Delta _{i}+g_{i} [\langle b\rangle +\langle b\rangle ^{\ast }] $ ($i=1, 2$%
) are the cavity detunings shifted by the radiation-pressure force. These
equations are coupled to each other and can be solved self-consistently.

Each operator of this system can be written as a sum of the steady-state
solution and its fluctuation with $a_{i}=\langle a_{i}\rangle +\delta a_{i}$
and $b=\langle b\rangle +\delta b$, where $\delta a_{i}$ are the
fluctuations of the cavities and $\delta b$ is that of the mechanical mode.
Neglecting the nonlinear terms in the radiation-pressure interaction in
Eqs.~(\ref{alpha1})-(\ref{beta}), we obtain a set of linear QLEs for the
fluctuation operators:
\begin{widetext}
\begin{eqnarray}
\delta \dot{a}_{1} &=&\left( -\gamma _{1}-i\Delta _{1}^{\prime }\right)
\delta a_{1}-iG_{1}\left( \delta b+\delta b^{\dag }\right) -iJ\delta
a_{2}+\varepsilon _{p}e^{i\left( \omega _{d}-\omega _{p}\right) t}+\xi _{1},
\\
\delta \dot{a}_{2} &=&\left( -\gamma _{2}-i\Delta _{2}^{\prime }\right)
\delta a_{2}-iG_{2}\left( \delta b+\delta b^{\dag }\right) -iJ\delta
a_{1}+\xi _{2}, \\
\delta \dot{b} &=&\left( -\gamma _{m}-i\omega _{m}\right) \delta b-i\left(
G_{1}\delta a_{1}^{\dag }+G_{1}^{\ast }\delta a_{1}\right) -i\left(
G_{2}\delta a_{2}^{\dag }+G_{2}^{\ast }\delta a_{2}\right)  \nonumber \\
&&+\varepsilon _{b}e^{-i\omega _{b}t}+\xi _{m},
\end{eqnarray}%
\end{widetext}
where $G_{i}=g_{i}\langle a_{i}\rangle $ ($i=1,2$) represent the
pump-enhanced linear optomechanical couplings.

In what follows, we fix $\omega _{b}=\omega _{p}-\omega _{d}$ in our scheme, i.e., the frequency of the mechanical drive is always equal to the frequency difference between the optical probe and pump fields. To solve the above QLEs, we transform all the operators to another rotating frame with $\delta a_{i}\rightarrow \delta
a_{i}e^{-i(\omega _{p}-\omega _{d})t}$, $\xi _{i}\rightarrow \xi
_{i}e^{-i(\omega _{p}-\omega _{d})t}$, $\delta b\rightarrow \delta
be^{-i\omega _{b}t}$, and $\xi _{m}\rightarrow \xi _{m}e^{-i\omega _{b}t}$.
In addition, we assume that the cavities are driven by the red-detuned pump fields and $\Delta _{i}^{\prime }\sim \omega _{m}$. In this case, by using the rotating-wave approximation, one can neglect the fast-oscillating counter-rotating terms and obtain the following linearized QLEs
\begin{eqnarray}
\delta \dot{a}_{1} &=&-\Gamma _{1}\delta a_{1}-iG_{1}\delta b-iJ\delta
a_{2}+\varepsilon _{p}+\xi _{1},  \label{eq:a1new} \\
\delta \dot{a}_{2} &=&-\Gamma _{2}\delta a_{2}-iG_{2}\delta b-iJ\delta
a_{1}+\xi _{2},  \label{eq:a2new} \\
\delta \dot{b} &=&-\Gamma _{m}\delta b-iG_{1}^{\ast }\delta
a_{1}-iG_{2}^{\ast }\delta a_{2}+\varepsilon _{b}+\xi _{m},  \label{eq:bnew}
\end{eqnarray}%
with $\Gamma _{i}=\gamma _{i}+i\Delta _{i}^{\prime \prime }$ and $\Gamma
_{m}=\gamma _{m}+i\Delta _{m}$. Here $\Delta _{i}^{\prime \prime }=\Delta
_{i}^{\prime }-(\omega _{p}-\omega _{d})$ and $\Delta _{m}=\omega
_{m}-(\omega _{p}-\omega _{d})$ are the detunings in the new rotating frame.
%
The optical response of the cavities to the probe field can be obtained by
solving the steady state of Eqs.~(\ref{eq:a1new})-(\ref{eq:bnew}). %
By setting $d\langle ...\rangle /dt=0$, we have
\begin{widetext}
\begin{eqnarray}
\left\langle \delta a_{1}\right\rangle &=&\frac{iG_{2}\varepsilon _{b}\left(
iJ\Gamma _{m}+G_{1}G_{2}^{\ast }\right) +\left( \Gamma _{2}\Gamma
_{m}+\left\vert G_{2}\right\vert ^{2}\right) \left( \varepsilon _{p}\Gamma
_{m}-iG_{1}\varepsilon _{b}\right) }{\left( \Gamma _{1}\Gamma
_{m}+\left\vert G_{1}\right\vert ^{2}\right) \left( \Gamma _{2}\Gamma
_{m}+\left\vert G_{2}\right\vert ^{2}\right) -\left( iJ\Gamma
_{m}+G_{1}G_{2}^{\ast }\right) \left( iJ\Gamma _{m}+G_{1}^{\ast
}G_{2}\right) },  \label{delta_1} \\
\left\langle \delta a_{2}\right\rangle &=&\frac{-\left( iJ\Gamma
_{m}+G_{1}^{\ast }G_{2}\right) \left( \varepsilon _{p}\Gamma
_{m}-iG_{1}\varepsilon _{b}\right) -iG_{2}\varepsilon _{b}\left( \Gamma
_{1}\Gamma _{m}+\left\vert G_{1}\right\vert ^{2}\right) }{\left( \Gamma
_{1}\Gamma _{m}+\left\vert G_{1}\right\vert ^{2}\right) \left( \Gamma
_{2}\Gamma _{m}+\left\vert G_{2}\right\vert ^{2}\right) -\left( iJ\Gamma
_{m}+G_{1}G_{2}^{\ast }\right) \left( iJ\Gamma _{m}+G_{1}^{\ast
}G_{2}\right) },  \label{delta_2} \\
\left\langle \delta b\right\rangle &=&\frac{J\Gamma _{m}\left( \varepsilon
_{b}J-\varepsilon _{p}G_{2}^{\ast }\right) +\Gamma _{2}\Gamma _{m}\left(
\varepsilon _{b}\Gamma _{1}-i\varepsilon _{p}G_{1}^{\ast }\right) }{\left(
\Gamma _{1}\Gamma _{m}+\left\vert G_{1}\right\vert ^{2}\right) \left( \Gamma
_{2}\Gamma _{m}+\left\vert G_{2}\right\vert ^{2}\right) -\left( iJ\Gamma
_{m}+G_{1}G_{2}^{\ast }\right) \left( iJ\Gamma _{m}+G_{1}^{\ast
}G_{2}\right) }.  \label{delta_b}
\end{eqnarray}
\end{widetext}

The cavity output fields $\langle \delta a_{i}^{out}\rangle $ ($i=1, 2$) can
be derived from the input-output theorem with
\begin{equation}
\left\langle \delta a_{i}^{out}\right\rangle +\left\langle \delta
a_{i}^{in}\right\rangle =\sqrt{2\gamma _{i}^{e}}\left\langle \delta
a_{i}\right\rangle,  \label{eq:inout}
\end{equation}%
where $\gamma _{i}^{e}$ represents the cavity loss related to coupling
between the cavity and the input (output) modes, and is part of the total
cavity loss rate $\gamma _{i}$ with $\gamma _{i}^{e}=\eta_{i}\gamma _{i}$
and $\eta _{i}\leq 1$. For simplicity of discussion, we focus on the case of
over-coupled cavities with $\eta _{i}\simeq 1$ and neglect cavity intrinsic
dissipation~\cite{Cai,Spillane, Tian2015}. With this assumption, $\langle
\delta a_{1}^{in}\rangle =\varepsilon _{p}/\sqrt{2\gamma _{1}^{e}}$, $%
\langle \delta a_{2}^{in}\rangle =0$.
The input field on the mechanical resonator can then be written in terms of the cavity input with $%
\langle \delta b^{in}\rangle =\sqrt{\gamma _{1}^{e}/\gamma _{m}}%
(ye^{i\varphi })\langle \delta a_{1}^{in}\rangle$.
The transmission coefficient that describes the dependence of the output
field of cavity $2$ on the input field $\langle \delta a_{1}^{in}\rangle$
can be defined as
\begin{equation}
t_{21}\equiv \partial \langle \delta a_{2}^{out}\rangle/\partial \langle
\delta a_{1}^{in}\rangle.
\end{equation}%
With Eqs.~(\ref{delta_2}) and (\ref{eq:inout}), we derive
\begin{widetext}
\begin{equation}
t_{21}=-2\sqrt{\gamma _{1}^{e}\gamma _{2}^{e}}\left[ \frac{\left( iJ\Gamma
_{m}+G_{1}^{\ast }G_{2}\right) \left( \Gamma _{m}-iG_{1}ye^{i\varphi
}\right) +iG_{2}ye^{i\varphi }\left( \Gamma _{1}\Gamma _{m}+\left\vert
G_{1}\right\vert ^{2}\right) }{\left( \Gamma _{1}\Gamma _{m}+\left\vert
G_{1}\right\vert ^{2}\right) \left( \Gamma _{2}\Gamma _{m}+\left\vert
G_{2}\right\vert ^{2}\right) -\left( iJ\Gamma _{m}+G_{1}G_{2}^{\ast }\right)
\left( iJ\Gamma _{m}+G_{1}^{\ast }G_{2}\right) }\right] ,  \label{t}
\end{equation}
\end{widetext}
where we have defined the amplitude of the mechanical drive through $\varepsilon_{b}/\varepsilon _{p} = ye^{i\varphi }$ ($y>0$).

\begin{figure*}[thb]
\centering
\includegraphics[bb=0 0 400 310, width=6cm, clip]{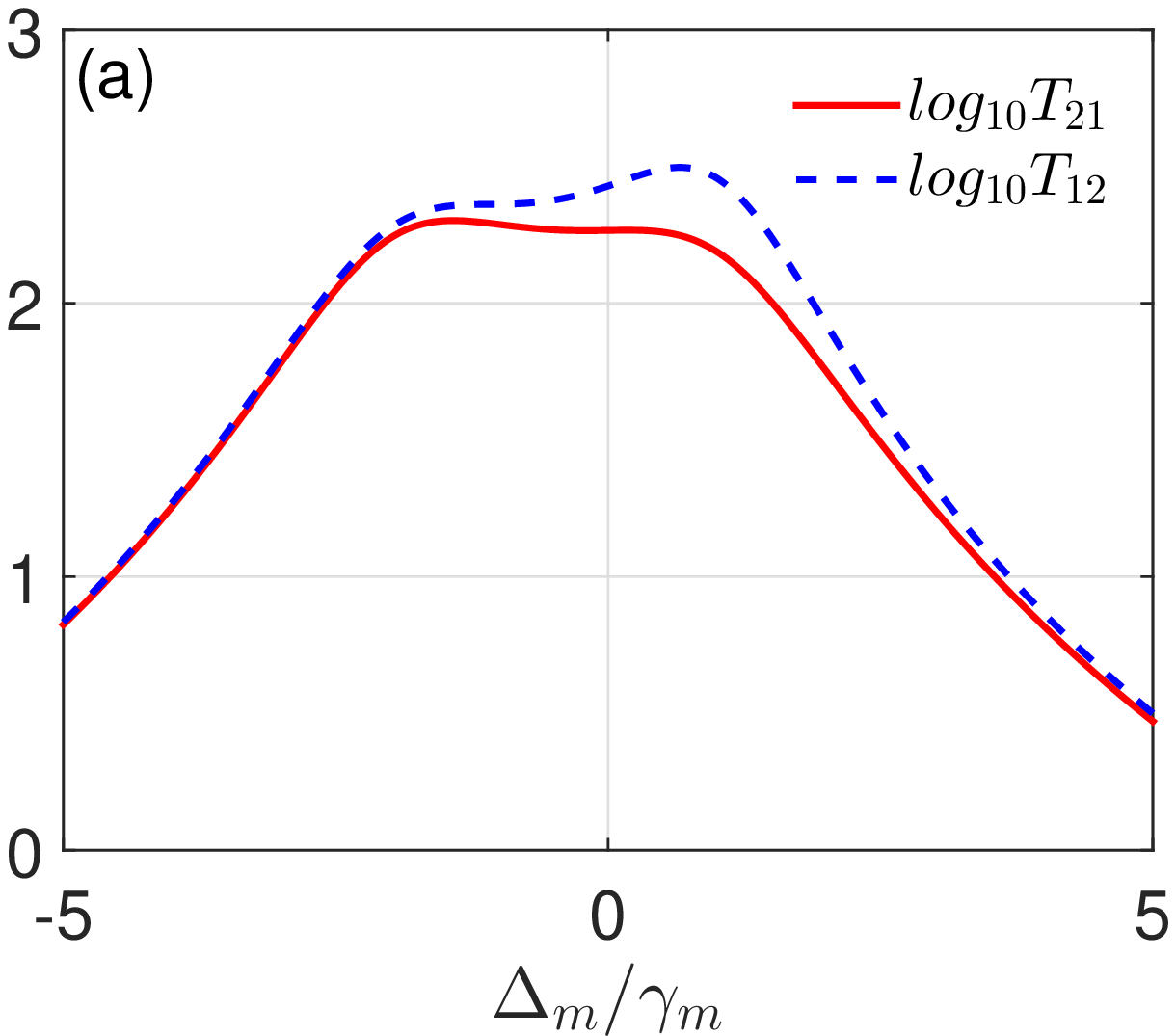} %
\includegraphics[bb=0 0 400 310, width=6cm, clip]{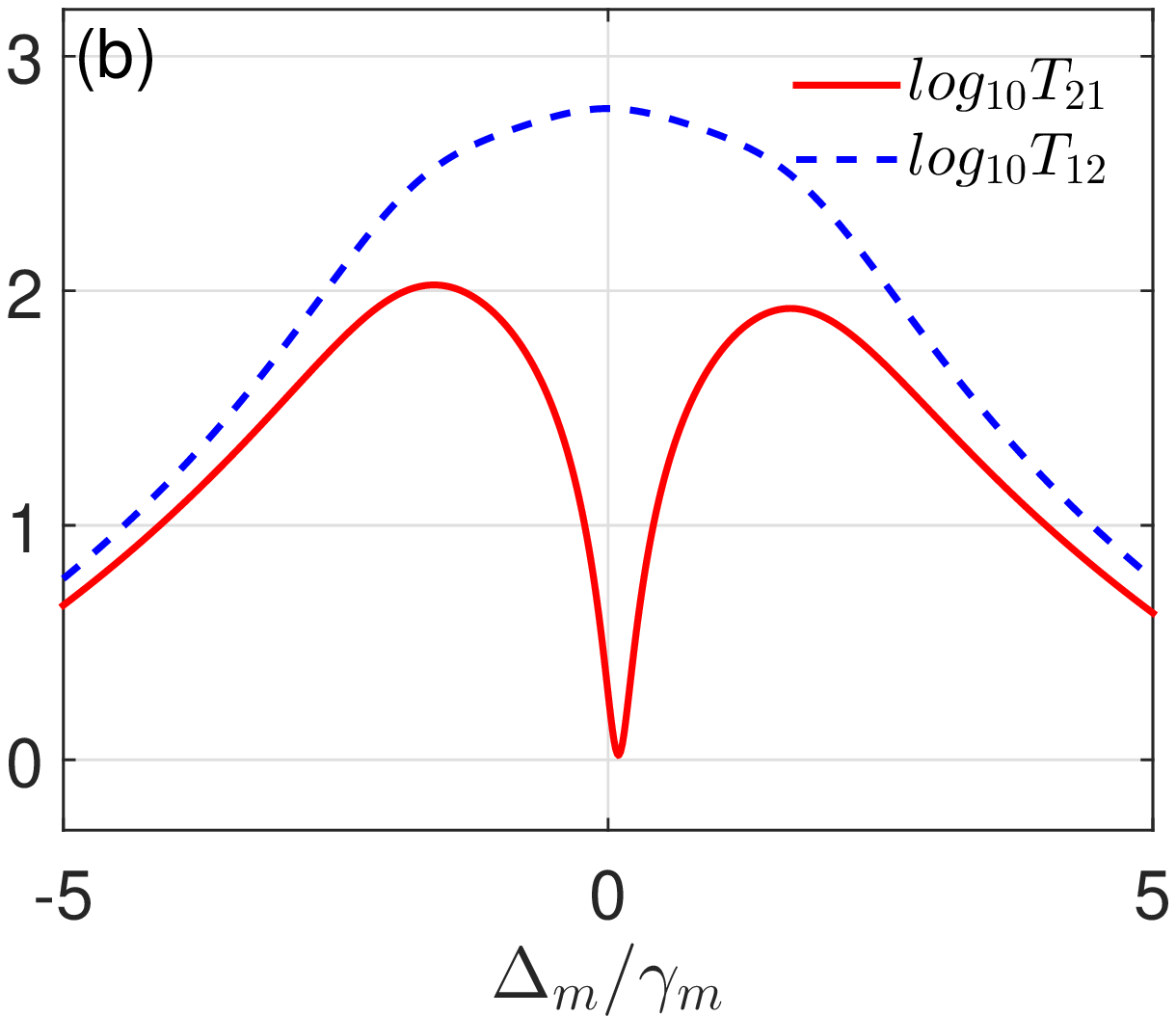} %
\includegraphics[bb=0 0 400 310, width=6cm, clip]{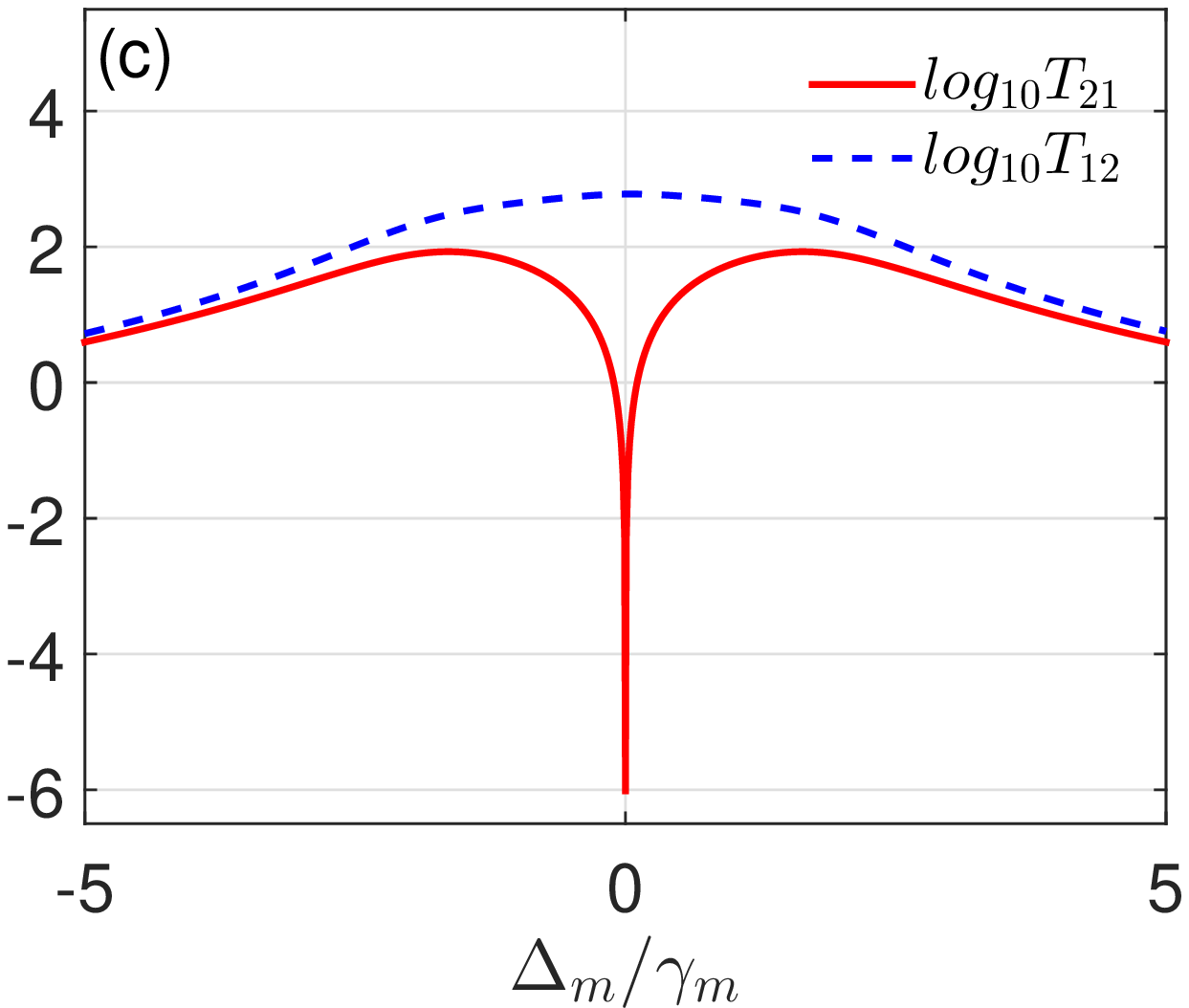}
\caption{The transmission probabilities $T_{21}$ and
$T_{12}$ versus $\Delta_m=\protect\omega
_{m}-(\protect\omega_{p}-\protect\omega_{d}) $ for
different values of
$\protect\theta $ and $\protect\varphi $: (a) $\protect\theta =0$,
$\protect\varphi =\protect\pi /2$; (b) $\protect\theta =\protect\pi /2$,
$\protect\varphi =0$; (c) $\protect\theta =\protect\pi /2$, $\protect\varphi
=\protect\pi /2$. Other parameters are $y=20$, $\protect\eta_{1,2} =1$,
$\protect\gamma _{1}=1.1\protect\gamma _{m}$, $\protect\gamma
_{2}=1.5\protect\gamma _{m}$, $G=\left\vert G_{1,2}\right\vert
=J=\protect\gamma _{m}$, and $\Delta _{1,2}^{\prime \prime }=\Delta _{m}$.}
\label{fig2}
\end{figure*}

Similarly, we can derive the transmission coefficient for a probe field applied to cavity $2$ from the right side (see the thin dashed arrow in Fig.~\ref{fig1}). In this case, we have $\langle \delta a_{1}^{in}\rangle=0$, $\langle \delta a_{2}^{in}\rangle= \varepsilon _{p}/\sqrt{2\gamma _{2}^{e}}$,  
and still fix $\omega_{b}=\omega_{p}-\omega_{d}$. Here the
transmission coefficient is defined as $t_{12}\equiv \partial \langle \delta
a_{1}^{out}\rangle/\partial \langle \delta a_{2}^{in}\rangle$. We derive
that
\begin{widetext}
\begin{equation}
t_{12}=-2\sqrt{\gamma _{1}^{e}\gamma _{2}^{e}}\left[ \frac{\left( iJ\Gamma
_{m}+G_{2}^{\ast }G_{1}\right) \left( \Gamma _{m}-iG_{2}ye^{i\varphi
}\right) +iG_{1}ye^{i\varphi }\left( \Gamma _{2}\Gamma _{m}+\left\vert
G_{2}\right\vert ^{2}\right) }{\left( \Gamma _{2}\Gamma _{m}+\left\vert
G_{2}\right\vert ^{2}\right) \left( \Gamma _{1}\Gamma _{m}+\left\vert
G_{1}\right\vert ^{2}\right) -\left( iJ\Gamma _{m}+G_{2}G_{1}^{\ast }\right)
\left( iJ\Gamma _{m}+G_{2}^{\ast }G_{1}\right) }\right] .  \label{t_21}
\end{equation}
\end{widetext}

This equation shows that the propagation of the optical probe field in the three-mode optomechanical system depends strongly on the interference between various paths of the probe field via the optical cavity with amplitude $\varepsilon_p$ and the frequency-matched mechanical drive with amplitude $\varepsilon_b$ via the optomechanical interaction. And the transmission is not symmetric between cavities $1$
and $2$.

\section{Directional amplification of optical probes}

\label{transmission}

In this section, we will study the transmission of optical probe and the
asymmetry in the transmission systematically. We will show that
amplification of optical probe fields can be directional.
Consider $G_{1}=G>0$ and $G_{2}=Ge^{i\theta }$ for simplicity of discussion.
The transmission coefficients can be rewritten as
\begin{widetext}
\begin{equation}
t_{21}=-2\sqrt{\gamma _{1}^{e}\gamma _{2}^{e}}\left[ \frac{\left( iJ\Gamma
_{m}+G^{2}e^{i\theta }\right) \left( \Gamma _{m}-iGye^{i\varphi }\right)
+iG\left( \Gamma _{1}\Gamma _{m}+G^{2}\right) ye^{i(\theta +\varphi )}}{%
\left( \Gamma _{1}\Gamma _{m}+G^{2}\right) \left( \Gamma _{2}\Gamma
_{m}+G^{2}\right) -\left( iJ\Gamma _{m}+G^{2}e^{-i\theta }\right) \left(
iJ\Gamma _{m}+G^{2}e^{i\theta }\right) }\right] ,  \label{tt21}
\end{equation}%
and
\begin{equation}
t_{12}=-2\sqrt{\gamma _{1}^{e}\gamma _{2}^{e}}\left[ \frac{\left( iJ\Gamma
_{m}+G^{2}e^{-i\theta }\right) \left( \Gamma _{m}-iGye^{i(\theta +\varphi
)}\right) +iG\left( \Gamma _{2}\Gamma _{m}+G^{2}\right) ye^{i\varphi }}{%
\left( \Gamma _{1}\Gamma _{m}+G^{2}\right) \left( \Gamma _{2}\Gamma
_{m}+G^{2}\right) -\left( iJ\Gamma _{m}+G^{2}e^{-i\theta }\right) \left(
iJ\Gamma _{m}+G^{2}e^{i\theta }\right) }\right] .  \label{tt12}
\end{equation}
\end{widetext}

When $\varepsilon _{b}=0$ ($y=0$), the model reduces to that studied in~\cite{XXW}, where the directional transmission of the probe field can be
achieved under optimal parameters. In such a scheme, the introduction of the
nontrivial phase $\theta $ breaks the time-reversal symmetry of this system
and results in nonreciprocal propagation of the probe field. %
In contrast, in the presence of the frequency-matched mechanical drive and in the absence of the second cavity ($J=0$ and $G_2=0$), the
system reduces to a standard two-mode optomechanical system. In this
case, it was shown that the presence of the mechanical drive $\varepsilon
_{b}$ leads to the amplification of the output field~\cite%
{OMAmplificationJia}. The amplification and enhancement in energy arise from
the phonon-photon parametric process in the presence of the
frequency-matched mechanical drive.

\begin{figure*}[tbh]
\centering
\includegraphics[bb=0 0 400 315, width=6cm, clip]{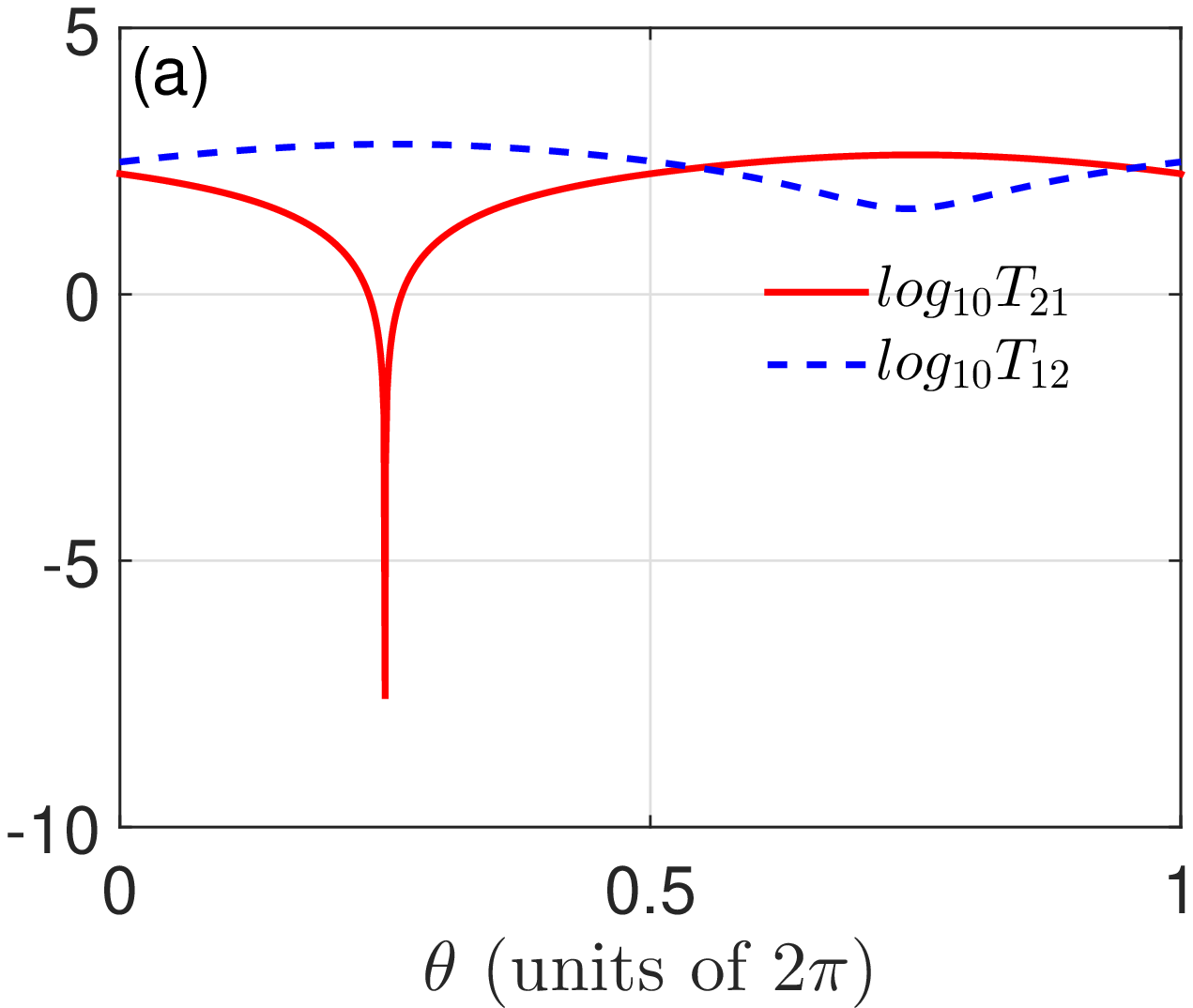} %
\includegraphics[bb=0 0 400 315, width=6cm, clip]{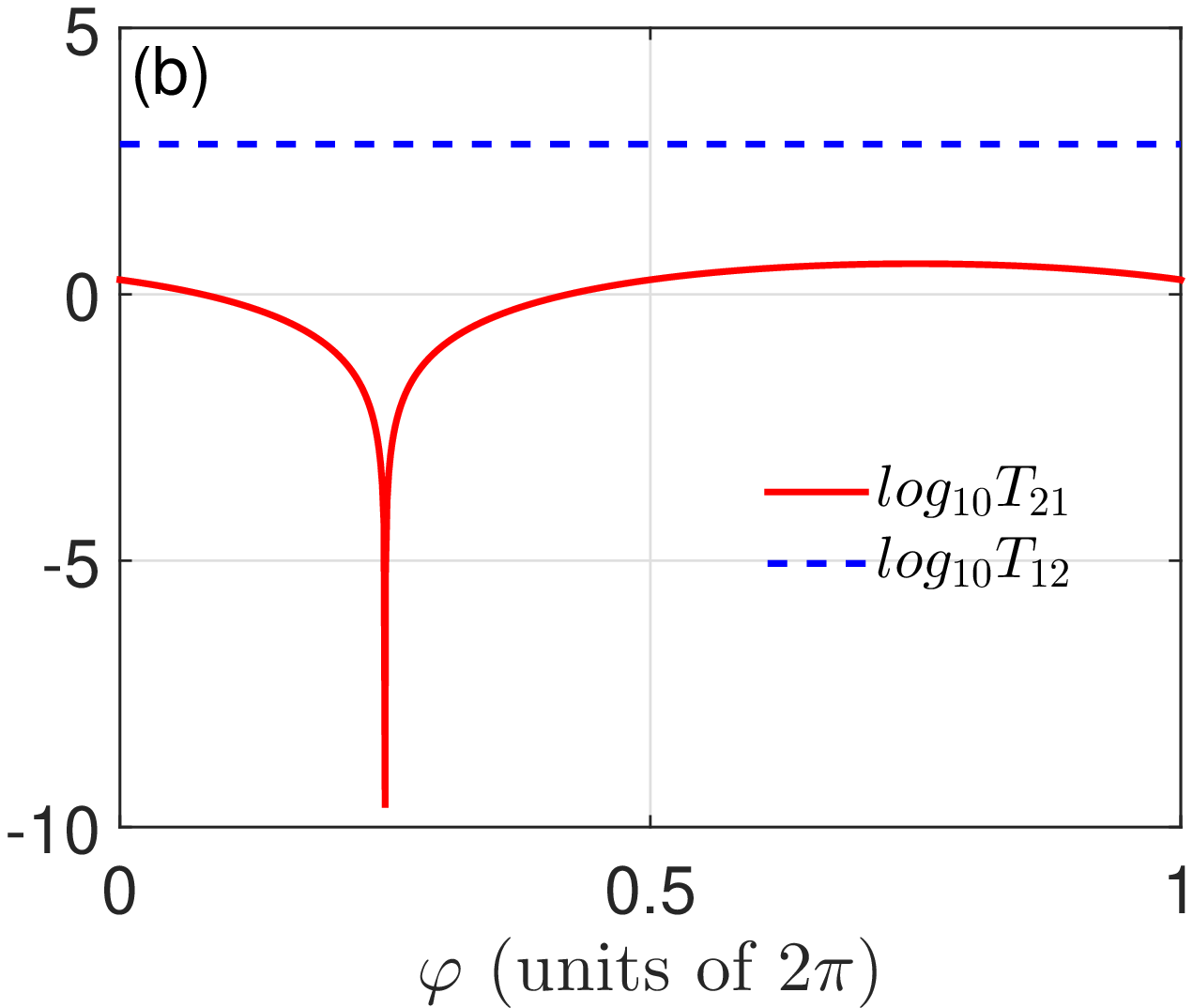}
\caption{Plot of the probability of transmission $T_{21}$
and $T_{12}$ as
functions of $\protect\theta $ and $\protect\varphi $,
respectively. (a) $\protect\varphi =\protect\pi /2$. (b) $\protect\theta
=\protect\pi /2$.
Other parameters are $y=20$, $\protect\eta _{1,2}=1$,
$G=\left\vert
G_{1,2}\right\vert =J=\protect\gamma _{m}$, $\Delta
_{m}=\Delta _{1,2}^{\prime \prime }=0$, $\protect\gamma
_{1}=1.1\protect\gamma _{m}$, and $\protect\gamma _{2}=1.5\protect\gamma
_{m}$. One can see that at certain optimal values of $\protect\theta $
and
$\protect\varphi $, e.g., $\protect\theta =\protect\pi /2$,
$\protect\varphi =\protect\pi /2$, $T_{12}\rightarrow 0$ and $T_{21}\gg 1$.
}
\label{fig3}
\end{figure*}

Now we study the effect of the frequency-matched mechanical drive $\varepsilon _{b}$ on the
propagation of the probe field in the three-mode optomechanical system in
the general case of $y\neq 0$ and $J\neq 0$. In Fig.~\ref{fig2}, we plot the probability of the transmission $%
T_{21}\equiv |t_{21}|^{2}$ and $T_{12}\equiv |t_{12}|^{2}$ as functions of $%
\Delta _{m}=\omega _{m}-(\omega _{p}-\omega _{d})$ at different values of
the phases $\theta $ and $\varphi $. We observe that in general, the
transmission of the probe field is asymmetric with $T_{21}\neq T_{12}$, and $T_{12}$ or $T_{21}$ can be much larger than $1$. This result indicates nonreciprocity with amplification of the optical probe field.
In particular, at certain optimal values of $\theta $ and $\varphi $, e.g., $\theta =\pi /2$, $\varphi =\pi /2$, $T_{21}\rightarrow 0$ and $T_{12}\gg 1$, as shown in Figs.~\ref{fig2}(c). The transmission from cavity $1$ to cavity $2$ is strongly amplified; whereas, the transmission on the opposite direction is suppressed. In this case, the amplification of the probe field results from phonon-photon parametric process due to the existence of the frequency-matched mechanical drive~\cite{OMAmplificationJia}.

We plot the probability of the transmission $T_{21}$ and $T_{12}$ as functions of $\theta $ and $\varphi $ in Fig. \ref{fig3}. It is also shown that the directional propagation can be achieved with $\theta =\pi /2$ in Fig.~\ref{fig3}(a) or $\varphi =\pi /2$ in Fig. \ref{fig3}(b). Note that, when $\theta =\pi /2$ with other parameters given in the caption of Fig. \ref{fig3}(b), the probability of
transmission $T_{12}$ is independent of $\varphi $, which can be given
through Eq. (\ref{tt12}).
%
\begin{figure}[h]
\centering
\includegraphics[bb=0 0 410 315, width=8cm, clip]{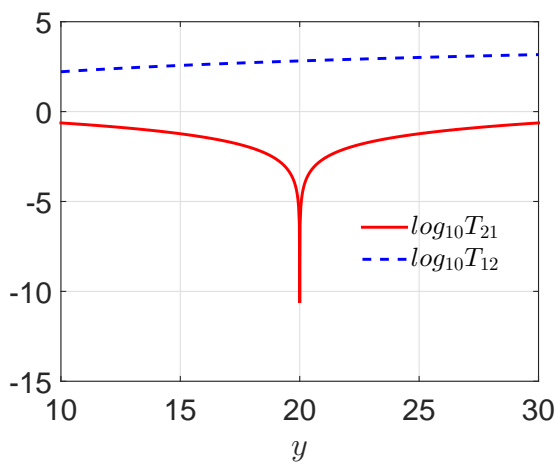}
\caption{The transmission probabilities $T_{21}$ and $T_{12}$ versus $y$.
Other parameters are $\protect\theta =\protect\pi /2$, $\protect\varphi =%
\protect\pi /2$, $G=\left\vert G_{1,2}\right\vert =J=\protect\gamma _{m}$, $%
\Delta _{m}=\Delta _{1,2}^{\prime \prime}=0$, $\protect\gamma _{1}=1.1%
\protect\gamma _{m}$, and $\protect\gamma
_{2}=1.5\protect\gamma _{m}$.}
\label{fig4}
\end{figure}

To further understand the effect of the frequency-matched mechanical drive on the transmission property of the probe field, we assume that the parameters $G=\left\vert G_{1,2}\right\vert =J=\gamma _{m}$, $%
\Delta _{m}=\Delta _{1,2}^{\prime \prime }=0$, $\theta =\pi /2$, and $%
\varphi =\pi /2$. Then the corresponding transmission coefficients $T_{12}$ and $T_{21}$ are simplified to be
\begin{eqnarray}
T_{21} &=&4\gamma _{1}\gamma _{2}\left( \frac{2\gamma _{m}\left( y+1\right)
-y\left( \gamma _{1}+\gamma _{m}\right) }{\left( \gamma _{1}+\gamma
_{m}\right) \left( \gamma _{2}+\gamma _{m}\right) }\right) ^{2},
\label{rt_1} \\
T_{12} &=&4y^{2}\frac{\gamma _{1}\gamma _{2}}{\left( \gamma _{1}+\gamma
_{m}\right) ^{2}}.  \label{rt_2}
\end{eqnarray}%
In the absence of the mechanical drive ($y\rightarrow
0$), the directional transmission of the probe field can occur with $T_{12}\rightarrow 0$ and $T_{21}> 0$ (particularly, $T_{21}=1$ at $\gamma _{1,2}=\gamma_{m} $) as shown in \cite{XXW}. On the contrary, in the presence of frequency-matched mechanical drive with $y=y_{c}\equiv 2\gamma_{m}/(\gamma _{1}-\gamma _{m})$ and $|y_{c}|\gg 1$, we have $%
T_{21}\rightarrow 0$ and $T_{12}\gg 1$. The directional amplification of the optical probe field can be observed due to the presence of the mechanical drive frequency-matched to the probe field, and the direction of the amplification is opposite to that in the case of directional transmission in \cite{XXW}. Strong amplification requires $|y_{c}|\gg 1$, i.e., the cavity damping rate $\gamma _{1}$ is approximately equal to the mechanical damping rate $\gamma _{m}$.

To study the role of the mechanical drive, we plot $T_{21}$ and $T_{12}$ as functions of $y$ in Fig.~\ref{fig4}. This plot clearly demonstrates that the propagation of the optical field is strongly amplified with $T_{12}\sim 600$ when the mechanical drive becomes large ($|y|\gg 1$). Meanwhile, when $y\sim y_{c}=20$ under the parameters given in the caption of Fig.~\ref{fig4}, the transmission in the opposite direction quickly drops with $T_{21}\rightarrow 0$.

\section{Conclusions}

\label{summary}

To conclude, we investigate the transmission of an optical probe field in a three-mode optomechanical system, where the mechanical resonator is subject to a mechanical drive with the driving frequency being equal to the frequency difference between the optical probe and pump fields. Under appropriate parameters, the directional amplification of the probe field resulting from the interference between different optical path and phonon-photon parametric process can be achieved. Amplification far exceeding unity can be achieved when the mechanical drive becomes strong. Such optomechanical setups could be used to switch and amplify weak probe signals in quantum networks.

\acknowledgments National Natural Science Foundation of China (with Grants. No.~11422437, No.~11505126, No.~11534002, No.~11421063, No.~U1530401); Postdoctoral Science Foundation of China (with Grant No. 2016M591055); PhD research startup foundation of Tianjin Normal University (with Grant. No. 52XB1415); National Science Foundation (NSF) (with Grants. No.~DMR-0956064 and No.~PHY-1720501); UC Multicampus-National Lab Collaborative Research and Training (with Grant No. LFR-17-477237).


\begin{thebibliography}{99}
\bibitem{Kippenberg} T. J. Kippenberg and K. J. Vahala, {\lq\lq}Cavity
optomechanics: back-action at the mesoscale,{\rq\rq} Science \textbf{321}(5893),
1172--1176 (2008).

\bibitem{Aspelmeyer1} M. Aspelmeyer, P. Meystre, and K. C. Schwab, {\lq\lq}Quantum
optomechanics,{\rq\rq} Phys. Today \textbf{65}(7), 29--35 (2012).

\bibitem{Meystre} P. Meystre, {\lq\lq}A short walk through quantum optomechanics,{\rq\rq}
Ann. Phys. (Berlin) \textbf{525}(3), 215--233 (2013).


\bibitem{Aspelmeyer3} M. Aspelmeyer, T. J. Kippenberg, and F. Marquardt,
{\lq\lq}Cavity optomechanics,{\rq\rq} Rev. Mod. Phys. \textbf{86}(4), 1391 (2014).

\bibitem{Vitali} D. Vitali, S. Gigan, A. Ferreira, H. R. B\"{o}hm, P.
Tombesi, A. Guerreiro, V.Vedral, A. Zeilinger, and M. Aspelmeyer,
{\lq\lq}Optomechanical entanglement between a movable mirror and a cavity field,{\rq\rq}
Phys. Rev. Lett. \textbf{98}(3), 030405 (2007).

\bibitem{Rugar} D. Rugar, R. Budakian, H. J. Mamin, and B. W. Chui, {\lq\lq}Single
spin detection by magnetic resonance force microscopy,{\rq\rq} Nature (London)
\textbf{430}(6997), 329--332 (2004).

\bibitem{Krause} A. G. Krause, M. Winger, T. D. Blasius, Q. Lin, and O.
Painter, {\lq\lq}A high-resolution microchip optomechanical accelerometer,{\rq\rq} Nat.
Photon. \textbf{6}(11), 768--772 (2012).

\bibitem{Regal} C. A. Regal, J. D. Teufel, and K. W. Lehnert, {\lq\lq}Measuring
nanomechanical motion with a microwave cavity interferometer,{\rq\rq} Nat. Phys.
\textbf{4}(7), 555--560 (2008).

\bibitem{Teufel} J. D. Teufel, T. Donner, M. A. Castellanos-Beltran, J. W.
Harlow, and K. W. Lehnert, {\lq\lq}Nanomechanical motion measured with an
imprecision below that at the standard quantum limit,{\rq\rq} Nat. Nanotechnol.
\textbf{4}(12), 820--823 (2009).

\bibitem{Forstner} S. Forstner, S. Prams, J. Knittel, E. D. van Ooijen, J.
D. Swaim, G. I. Harris, A. Szorkovszky, W. P. Bowen, and H.
Rubinsztein-Dunlop, {\lq\lq}Cavity optomechanical magnetometer,{\rq\rq} Phys. Rev. Lett.
\textbf{108}(12), 120801 (2012).

\bibitem{Xu} X. Xu and J. M. Taylor, {\lq\lq}Squeezing in a coupled two-mode
optomechanical system for force sensing below the standard quantum limit,{\rq\rq}
Phys. Rev. A \textbf{90}(4), 043848 (2014).

\bibitem{Arvanitaki} A. Arvanitaki and A. A. Geraci, {\lq\lq}Detecting
high-frequency gravitational waves with optically levitated sensors,{\rq\rq} Phys.
Rev. Lett. \textbf{110}(7), 071105 (2013).

\bibitem{Mancini} S. Mancini, D. Vitali, and P. Tombesi, {\lq\lq}Scheme for
teleportation of quantum states onto a mechanical resonator,{\rq\rq} Phys. Rev.
Lett. \textbf{90}(13), 137901 (2003).

\bibitem{Barzanjeh} Sh. Barzanjeh, S. Guha, C. Weedbrook, D. Vitali, J. H.
Shapiro, and S. Pirandola, {\lq\lq}Microwave quantum illumination,{\rq\rq} Phys. Rev.
Lett. \textbf{114}(8), 080503 (2015).

\bibitem{OMIT1} G. S. Agarwal and S. Huang, {\lq\lq}Electromagnetically induced
transparency in mechanical effects of light,{\rq\rq} Phys. Rev. A \textbf{81}(4),
041803 (2010).

\bibitem{OMIT2} S. Weis, R. Rivi\`{e}re, S. Del\'{e}glise, E. Gavartin, O.
Arcizet, A. Schliesser, and T. J. Kippenberg, {\lq\lq}Optomechanically induced
transparency,{\rq\rq} Science \textbf{330}(6010), 1520--1523 (2010).

\bibitem{OMIT3} J. D. Teufel, D. Li, M. S. Allman, K. Cicak, A. J. Sirois,
J. D. Whittaker, and R. W. Simmonds, {\lq\lq}Circuit cavity electromechanics in the
strong-coupling regime,{\rq\rq} Nature (London), \textbf{471}(7337), 204--208 (2011).

\bibitem{OMIT4} A. H. Safavi-Naeini, T. P. M. Alegre, J. Chan, M.
Eichenfield, M. Winger, Q. Lin, J. T. Hill, D. E. Chang, and O. Painter,
{\lq\lq}Electromagnetically induced transparency and slow light with
optomechanics,{\rq\rq} Nature (London) \textbf{472}(7341), 69--73 (2011).

\bibitem{OMIT5} M. Karuza, C. Biancofiore, M. Bawaj, C. Molinelli, M.
Galassi, R. Natali, P. Tombesi, G. Di Giuseppe, and D. Vitali,
{\lq\lq}Optomechanically induced transparency in a membrane-in-the-middle setup at
room temperature,{\rq\rq} Phys. Rev. A \textbf{88}(1), 013804 (2013).

\bibitem{OMslowlight1} X. Zhou, F. Hocke, A. Schliesser, A. Marx, H. Huebl,
R. Gross, and T. J. Kippenberg, {\lq\lq}Slowing, advancing and switching of
microwave signals using circuit nanoelectromechanics,{\rq\rq} Nat. Phys. \textbf{9}%
(3), 179--184 (2013).

\bibitem{OMslowlight2} D. E. Chang, A. H. Safavi-Naeini, M. Hafezi, and O.
Painter, {\lq\lq}Slowing and stopping light using an optomechanical crystal array,{\rq\rq}
New J. Phys. \textbf{13}(2), 023003 (2011).

\bibitem{OMIA1} F. Hocke, X. Zhou, A. Schliesser, T. J. Kippenberg, H.
Huebl, and R. Gross, {\lq\lq}Electromechanically induced absorption in a circuit
nano-electromechanical system,{\rq\rq} New J. Phys. \textbf{14}(12), 123037 (2012).

\bibitem{OMIA2} K. Qu and G. S. Agarwal, {\lq\lq}Phonon-mediated
electromagnetically induced absorption in hybrid opto-electromechanical
systems,{\rq\rq} Phys. Rev. A \textbf{87}(3), 031802 (2013).

\bibitem{OMAmplification1} F. Massel, T. T. Heikkil\"{a}, J.-M.
Pirkkalainen, S. U. Cho, H. Saloniemi, P. J. Hakonen, and M. A. Sillanp\"{a}%
\"{a}, {\lq\lq}Microwave amplification with nanomechanical resonators,{\rq\rq} Nature
(London) \textbf{480}(7377), 351--354 (2011).

\bibitem{OMAmplification2} A. Metelmann and A. A. Clerk, {\lq\lq}Quantum-limited
amplification via reservoir engineering,{\rq\rq} Phys. Rev. Lett. \textbf{112}(13),
133904 (2014).


\bibitem{OMAmplificationJia} W. Z. Jia, L. F. Wei, Y. Li, and Y. X. Liu,
{\lq\lq}Phase-dependent optical response properties in an optomechanical system by
coherently driving the mechanical resonator,{\rq\rq} Phys. Rev. A \textbf{91}(4),
043843 (2015).

\bibitem{OMAmplificationXu} X.-W. Xu and Y. Li, {\lq\lq}Controllable optical output
fields from an optomechanical system with mechanical driving,{\rq\rq} Phys. Rev. A
\textbf{92}(2), 023855 (2015).

\bibitem{OMAmplificationSSi} L.-G. Si, H. Xiong, M. S. Zubairy, and Y. Wu,
{\lq\lq}Optomechanically induced opacity and amplification in a quadratically
coupled optomechanical system,{\rq\rq} Phys. Rev. A \textbf{95}, 033803 (2017).

\bibitem{Haus} H. A. Haus, \textit{Waves and Fields in Optoelectronics}
(Prentice-Hall, Englewood Cliffs, NJ, 1984).

\bibitem{Faraday} L. Bi, J. Hu, P. Jiang, D. H. Kim, G. F. Dionne, L. C.
Kimerling, and C. A. Ross, {\lq\lq}On-chip optical isolation in monolithically
integrated non-reciprocal optical resonators,{\rq\rq} Nat. Photon. \textbf{5}(12),
758--762 (2011).

\bibitem{ParametricModulation1} Z. Yu and S. Fan, {\lq\lq}Complete optical
isolation created by indirect interband photonic transitions,{\rq\rq} Nat. Photon.
\textbf{3}(2), 91--94 (2009).

\bibitem{ParametricModulation2} D. W. Wang, H. T. Zhou, M. J. Guo, J. X.
Zhang, J. Evers, and S. Y. Zhu, {\lq\lq}Optical diode made from a moving photonic
crystal,{\rq\rq} Phys. Rev. Lett. \textbf{110}(9), 093901 (2013).

\bibitem{ParametricModulation3} S. A. R. Horsley, J.-H. Wu, M. Artoni, and
G. C. La Rocca, {\lq\lq}Optical nonreciprocity of cold atom Bragg mirrors in
motion,{\rq\rq} Phys. Rev. Lett. \textbf{110}(22), 223602 (2013).

\bibitem{ParametricModulation4} N. A. Estep, D. L. Sounas, J. Soric, and A.
Al\`{u}, {\lq\lq}Magnetic-free non-reciprocity and isolation based on
parametrically modulated coupled-resonator loops,{\rq\rq} Nat. Phys. \textbf{10}%
(12), 923--927 (2014).

\bibitem{Chang} L. Chang, X. Jiang, S. Hua, C. Yang, J. Wen, L. Jiang, G.
Li, G. Wang, and M. Xiao, {\lq\lq}Parity-time symmetry and variable optical
isolation in active-passive-coupled microresonators,{\rq\rq} Nat. Photon. \textbf{8}%
(7), 524--529 (2014).

\bibitem{Tang} X. Guo, C.-L. Zou, H. Jung, and H. X. Tang, {\lq\lq}On-chip strong
coupling and efficient frequency conversion between telecom and visible
optical modes,{\rq\rq} Phys. Rev. Lett. \textbf{117}(12), 123902 (2016).

\bibitem{Chiral} I. S\"{o}llner, S. Mahmoodian, S. L. Hansen, L. Midolo, A.
Javadi, G. Kir\v{s}ansk\.{e}, T. Pregnolato, H. El-Ella, E. H. Lee, J. D.
Song, S\o ren Stobbe, and P. Lodahl , {\lq\lq}Deterministic photon--emitter
coupling in chiral photonic circuits,{\rq\rq} Nat. Nanotechnol. \textbf{10}(9),
775--778 (2015).

\bibitem{Lecocq} F. Lecocq, L. Ranzani, G. A. Peterson, K. Cicak, R. W. Simmonds, J. D. Teufel, and J. Aumentado, {\lq\lq}Nonreciprocal microwave signal processing with a field-programmable Josephson amplifier,{\rq\rq} Phys. Rev. Applied \textbf{7}(2), 024028 (2017).

\bibitem{Manipatruni} S. Manipatruni, J. T. Robinson, and M. Lipson,
{\lq\lq}Optical nonreciprocity in optomechanical structures,{\rq\rq} Phys. Rev. Lett.
\textbf{102}(21), 213903 (2009).

\bibitem{Hafezi} M. Hafezi and P. Rabl, {\lq\lq}Optomechanically induced
non-reciprocity in microring resonators,{\rq\rq} Opt. Express \textbf{20}(7),
7672--7684 (2012).

\bibitem{shen} Z. Shen, Y.-L. Zhang, Y. Chen, C.-L. Zou, Y.-F. Xiao, X.-B.
Zou, F.-W. Sun, G.-C. Guo, and C.-H. Dong, {\lq\lq}Experimental realization of
optomechanically induced non-reciprocity,{\rq\rq} Nat. Photon. \textbf{10}(10),
657--661 (2016).

\bibitem{Kim} J. Kim, M. C. Kuzyk, K. Han, H. Wang, and G. Bahl,
{\lq\lq}Non-reciprocal Brillouin scattering induced transparency,{\rq\rq} Nat. Phys.
\textbf{11}(3), 275--280 (2015).


\bibitem{Fang} K. Fang, J. Luo, A. Metelmann, M. H. Matheny, F. Marquardt,
A. A. Clerk, and O. Painter, {\lq\lq}Generalized nonreciprocity in an
optomechanical circuit via synthetic magnetism and reservoir engineering,{\rq\rq}
Nat. Phys. \textbf{13}, 465 (2017).


\bibitem{XXW} X. W. Xu and Y. Li, {\lq\lq}Optical nonreciprocity and optomechanical
circulator in three-mode optomechanical systems,{\rq\rq} Phys. Rev. A \textbf{91}%
(5), 053854 (2015).

\bibitem{XXW2} X. W. Xu, Y. Li, A. X. Chen, and Y. X. Liu, {\lq\lq}Nonreciprocal
conversion between microwave and optical photons in electro-optomechanical
systems,{\rq\rq} Phys. Rev. A \textbf{93}(2), 023827 (2016).

\bibitem{Tian} L. Tian and Z. Li, {\lq\lq}Nonreciprocal state conversion between
microwave and optical Photons,{\rq\rq} arXiv: 1610.09556 (2016).

\bibitem{Metelmann}
A. Metelmann, and A. A. Clerk, {\lq\lq}Nonreciprocal photon
transmission and amplification via reservoir engineering,{\rq\rq} Phys. Rev. X
\textbf{5}(2), 021025 (2015).

\bibitem{YLZhang}
Y. L. Zhang, C. H. Dong, C. L. Zou, X. B. Zou, Y. D. Wang,
and G. C. Guo, {\lq\lq}Optomechanical devices based on traveling-wave
microresonators,{\rq\rq} Phys. Rev. A \textbf{95}(4), 043815 (2017).

\bibitem{Ranzani} L. Ranzani, and J. Aumentado, {\lq\lq}Graph-based
analysis of nonreciprocity in coupled-mode systems,{\rq\rq} New J. Physics,
\textbf{17}(2), 023024 (2015).

\bibitem{Peterson} G. A. Peterson, F. Lecocq, K. Cicak, R. W. Simmonds, J.
Aumentado, and J. D. Teufel, {\lq\lq}Demonstration of efficient nonreciprocity in a
microwave optomechanical circuit,{\rq\rq} arXiv: 1703.05269.

\bibitem{Malz} D. Malz, L. D. Toth, N. R. Bernier, A. K. Feofanov, T. J.
Kippenberg, and A. Nunnenkamp, {\lq\lq}Quantum-limited directional amplifiers with
optomechanics,{\rq\rq} arXiv: 1705.00436.

\bibitem{Ruesink} F. Ruesink, M. A. Miri, A. Al\`{u}, and E. Verhagen,
{\lq\lq}Nonreciprocity and magnetic-free isolation based on optomechanical
interactions,{\rq\rq} Nat. Commun. \textbf{7}, 13662 (2016).

\bibitem{Mohammad} M. A. Miri, F. Ruesink, E. Verhagen, and
A. Al\`{u}, {\lq\lq}Fundamentals of optical non-reciprocity based on optomechanical
coupling,{\rq\rq} Phys. Rev. Applied \textbf{7}(6), 064014 (2017).

\bibitem{Bernier} N. R. Bernier, L. D. T\'{o}th, A. Koottandavida, M.
Ioannou, D. Malz, A. Nunnenkamp, A. K. Feofanov, and T. J. Kippenberg,
{\lq\lq}Nonreciprocal reconfigurable microwave optomechanical circuit,{\rq\rq} arXiv:
1612.08223 (2016).


\bibitem{Rugar1} D. Rugar and P. Gr\"{u}tter, {\lq\lq}Mechanical parametric amplification and thermomechanical noise squeezing,{\rq\rq} Phys. Rev. Lett. \textbf{67}(6), 699 (1991).
\bibitem{Okamoto} H. Okamoto, A. Gourgout, C. Y. Chang, K. Onomitsu, I. Mahboob, E. Y. Chang, and H. Yamaguchi, {\lq\lq}Coherent phonon manipulation in coupled mechanical resonators,{\rq\rq} Nat. Phys. \textbf{9}(8), 480--484 (2013).
\bibitem{Faust} T. Faust, J. Rieger, M. J. Seitner, J. P. Kotthaus, and E. M. Weig, {\lq\lq}Coherent control of a classical nanomechanical two-level system,{\rq\rq} Nat. Phys. \textbf{9}(8), 485--488 (2013).
\bibitem{Fu} H. Fu, Z. Gong, T. Mao, C. Sun, S. Yi, Y. Li, and G. Cao, {\lq\lq}Classical analog of St\"{u}ckelberg interferometry in a two-coupled-cantilever¨Cbased optomechanical system,{\rq\rq} Phys. Rev. A \textbf{94}(4), 043855 (2016).


\bibitem{Cai} M. Cai, O. J. Painter, and K. J. Vahala, {\lq\lq}Observation of
critical coupling in a fiber taper to a silica-microsphere
whispering-gallery mode system,{\rq\rq} Phys. Rev. Lett. \textbf{85}(1), 74 (2000).

\bibitem{Spillane} S. M. Spillane, T. J. Kippenberg, O. J. Painter, and K.
J. Vahala, {\lq\lq}Ideality in a fiber-taper-coupled microresonator system for application to cavity quantum electrodynamics,{\rq\rq} Phys. Rev. Lett. \textbf{91}(4), 043902 (2003).

\bibitem{Tian2015} L. Tian, {\lq\lq}Optoelectromechanical transducer: Reversible
conversion between microwave and optical photons,{\rq\rq} Ann. Phys. (Berlin)
\textbf{527}(1-2), 1--14 (2015).
\end{thebibliography}
\end{document}